\documentclass[a4paper,10pt,abstracton,openbib,final]{scrartcl}
\usepackage[a4paper,includehead,top=10mm,bottom=20mm,right=20mm,left=20mm]{geometry}
\usepackage[final]{graphicx}
\usepackage{graphics}
\usepackage{subfig}
\usepackage{mathrsfs}
\usepackage{amsmath, amsthm, amsxtra}
\usepackage{accents}
\usepackage{amsfonts}
\usepackage{amssymb}
\usepackage{stmaryrd}
\usepackage{lmodern}
\usepackage{textcomp}
\usepackage{booktabs}
\usepackage{eucal}
\usepackage{bm}
\usepackage{subeqnarray}
\usepackage{graphicx} 
\usepackage{upgreek}
\DeclareMathAlphabet{\mathscr}{OT1}{pzc}{m}{it}

\usepackage{hyperref}

\renewcommand{\t}[1]{\bm{#1}}

\newcommand{\p}{\partial}
\renewcommand{\d}{\, \mathrm d }

\newcommand{\del}{\updelta}
\newcommand{\dd}[2]{\frac{\d #1}{\d #2}}
\newcommand{\pd}[2]{\frac{\p #1}{\p #2}}
\newcommand{\comma}{\ , \quad }

\newcommand{\grad}{\nabla}
\newcommand{\D}{\Delta }
\newcommand{\Dt}[1]{#1^{\scalebox{0.5}{\textbullet} } }

\newcommand{\Om}{\Omega}
\renewcommand{\v}{\mathrm v}

\newcommand{\begeq}{\begin{equation}\begin{gathered}}
\newcommand{\eqend}{\end{gathered}\end{equation}}

\newcommand{\jl}{\llbracket}
\newcommand{\jr}{\rrbracket}
\newcommand{\Ref}{\text{\tiny ref.}}

\title{\huge An accurate finite element method for the numerical solution of isothermal and incompressible flow of viscous fluid}
\author{
B. Emek Abali
\thanks{Corresponding author, email: bilenemek@abali.org}
\thanks{Technische Universit\"at Berlin, Institute of Mechanics}
}
\date{}

\begin{document}
\maketitle

\begin{abstract}
Despite its numerical challenges, finite element method is used to compute viscous fluid flow. A consensus on the cause of numerical problems has been reached; however, general algorithms---allowing a robust and accurate simulation for any process---are still missing. Either a very high computational cost is necessary for a direct numerical solution (DNS) or some limiting procedure is used by adding artificial dissipation to the system. These stabilization methods are useful; however, they are often applied relative to the element size such that a local monotonous convergence is challenging to acquire. We need a computational strategy for solving viscous fluid flow using solely the balance equations. In this work, we present a general procedure solving fluid mechanics problems without use of any stabilization or splitting schemes. Hence, its generalization to multiphysics applications is straightforward. We discuss emerging numerical problems and present the methodology rigorously. Implementation is achieved by using open-source packages and the accuracy as well as the robustness is demonstrated by comparing results to the closed-form solutions and also by solving well-known benchmarking problems.
\end{abstract}

\paragraph{Keywords:} Finite element method, Fluid dynamics, Computation, Viscous fluid flow

\section{Introduction}

Isothermal flow of viscous fluid is modeled in Cartesian coordinates by using the balance equations of mass and linear momentum:
\begeq \label{balances.local}
\pd{\rho}{t} + \pd{v_i \rho }{x_i} = 0 \comma
\pd{\rho v_j}{t} + \pd{}{x_i} \Big( v_i \rho v_j - \sigma_{ij} \Big) =  \rho g_j \ ,
\eqend
respectively, where $\rho$ denotes the mass density, $v_i$ the velocity, $\sigma_{ij}$ the non-convective flux term (\textsc{Cauchy}'s stress), $g_i$ the specific supply (gravitational forces); here and henceforth we apply \textsc{Einstein}'s summation convention to repeated indices. In the case of \textsc{Newton}ian fluids such as water, oil, or alcohol, a linear relation for stress furnishes the governing equations with sufficient accuracy. This linear relation is sometimes called the \textsc{Navier--Stokes} equation:\footnote{In the literature, often the law of motion, i.e., the material equation inserted into the balance equation is called \textsc{Navier--Stokes} equation. \textsc{Navier} did use \textsc{Lagrange}an method for derivation of the law of motion; however, \textsc{Stokes} used---as we do it herein as well---the method introduced by \textsc{Cauchy} for separating material equation from the balance equation, see \cite{darrigol2002between} for historical remarks.} 
\begeq \label{constitutive}
\sigma_{ij} = ( -p + \lambda d_{kk} ) \delta_{ij} + 2 \mu d_{ij} \comma 
d_{ij} = \frac12 \Big( \pd{v_i}{x_j} + \pd{v_j}{x_i} \Big) \ ,
\eqend
with the material constants $\lambda$, $\mu$; and a new parameter called (hydrostatic) pressure, $p$. Consider an incompressible flow in a control volume initially filled with homogeneous water. Thus, the mass density remains constant in space and time; from the mass balance in Eq.\,\eqref{balances.local}$_1$, we obtain,
\begeq \label{mass.incomp.local}
\pd{v_i}{x_i} = 0 \ ,
\eqend
which is used for computing the pressure $p$. For an incompressible flow---as seen from Eqs.\,\eqref{mass.incomp.local}, \eqref{constitutive}---the mechanical pressure $-\frac13 \sigma_{ii}$ becomes identical to the hydrostatic pressure $p$ such that we handle $p$ as the pressure generated in a pump. Velocity and pressure fields have to satisfy Eqs.\,\eqref{balances.local}$_2$ and \eqref{mass.incomp.local}. 

In analytical mechanics, the aforementioned equations are fulfilled locally (in every \textit{infinitesimal} point in space). For a computation we discretize the space, herein by using the finite element method (FEM). Within each element, the analytical functions for the unknowns $v_i$ and $p$ are represented by form (shape) functions with a local support, i.e., by means of a discrete element. The shape functions are not smooth, they belong to $C^n$ with a finite $n$. In other words, the unknowns are finitely differentiable and depending on the governing equations and constitutive relations---from the mathematical analysis in \cite{babuska}---we know that the correct choice of the form functions for velocity and pressure is of paramount importance for a robust computation. This so-called \textsc{Ladyzhenskaya--Babuska--Brezzi} (inf-sup compatibility) condition (LBB condition) tells us how to adjust the shape functions of velocity and pressure in the case of an isothermal and incompressible flow. Special elements like the \textsc{Taylor--Hood} element \cite{taylor1973numerical} or a mixed element with bubble functions \cite{arnold1984stable} are often used for the isothermal and incompressible fluid flow problems. If one wants to include temperature deviation and electromagnetism into the computation, we fail to know the corresponding LBB condition for all shape functions (velocity, pressure, temperature, and electromagnetic fields). For practical purposes, a robust computation without exploiting the LBB condition is useful for a straightforward extension to multiphysics applications. This aspect is the main motivation of this work.

\subsection{Computational fluid dynamics (CFD)}
There are several methods for solving the aforementioned equations numerically. As already indicated, we concentrate on the finite element method in this work. Within a finite element, the governing equations are satisfied globally (over the domain of the element). There exists a general assumption that we can use the same local governing equations holding globally in finite elements; however, this strategy leads to several numerical problems and to various proposals in \cite{book1975, hughes83, hughes, tezduyar, silvester1990, franca1995, blank1999, brezzi2001, gravemeier2006}, for a review of such suggestions see \cite{morton}. These so-called stabilization methods introduce a numerical parameter depending on the underlying mesh. There are several successful implementations as in \cite{kuzmin2002, encyc1, encyc2, donea, hoffman_johnson, lohner2008, forster2009, arzani2016}. From a practical point of view, such a stabilization method is very useful; however, introducing numerical parameters may be challenging as they need to be tuned depending on the application, see \cite{barth2004} for a discussion about this issue leading to methods being conditionally stable; see references in \cite{bowers2010} about accuracy problems of different techniques. Indeed, numerical strategies exist for performing accurate simulations without numerical parameters. By using small elements---smaller than the characteristic length scale, for example \textsc{Kolmogorov} scale---we can overcome the numerical problems. This direct numerical simulation (DNS) is accurate and robust as demonstrated in \cite{eggels1994fully, hirsch2007numerical, varghese2007direct, lee2015direct, tryggvason20162, hoffman2015fenics, nguyen2018high}; however, it is often not feasible without access to super-computers. There is another class of so-called splitting or projection methods such as \textsc{Chorin}'s method or its derivatives as in \cite{chorin1969, temam1969a, temam1969b, goda1979, bell1989, zienkiewicz1995, brown2001}. The finite volume method (FVM) is often used in the computational fluid dynamics (CFD) as it is accurate and stable \cite{anderson1995computational, jasak1996error, versteeg2007introduction, wang2013high}. These methods are reliable; but it is challenging to adopt a splitting method or FVM in multiphysics. By using the FEM for viscous fluid flows, we intend to design a computation strategy by employing only physical (measurable) parameters in such a way that the strategy shows a local monotonous convergence as expected from the FEM. 

The briefly mentioned problems are well known in the literature such that various new computation methods are suggested for viscous flow problems. The (numerical) parameter free approach in \cite{dohrmann2004} shows 2D and 3D results for stationary viscous flows without the nonlinear convection term (often called \textsc{Stokes}'s problem). Based on this idea an under-integrated mass matrix is used in \cite{he2008, li2008} to perform simulations without stabilization terms in 2D. Several mesh-dependent stabilization terms and their connections to mesh-independent stabilization methods are investigated in \cite{burman2008}. In \cite{olshanskii2010} vorticity is used instead of velocity such that a new kind of splitting scheme is proposed for solving 3D problems. In \cite{011} balance equations of mass, momentum, angular momentum, and energy are used for performing 3D computations without numerical parameters; however, the method already uses the energy equation such that generalization for the non-isothermal case seems to be quite difficult. In \cite{palha2017} vorticity is introduced as an independent term ensuring that the balance of moment of momentum is satisfied, in 2D numerical solutions are performed without necessitating any (numerical) parameters. In \cite{charnyi2017} different strategies are performed for establishing 3D simulations. They are all based on writing the nonlinear convection term in a different (mathematically equivalent) form. In \cite{fu2018} a gradient-velocity-pressure formulation is suggested to solve 2D numerical experiments.

\subsection{Scope of this work}
In this work, we discuss a special yet general case, namely an isothermal and incompressible flow. For understanding the numerical problems, often, the pressure related numerical problems and velocity related numerical problems are studied separately. We use one balance equation for calculating the pressure and another balance equation for calculating the velocity. These balance equations are coupled such that we fail to uniquely identify the appropriate balance equation to be used for pressure or velocity. Therefore, more robust numerical strategies use both balance equations for both of the unknowns, this approach has already been undertaken in \textsc{Chorin}'s method and then extensively exploited by the pressure stabilized \textsc{Petrov--Galerkin} method (PSPG). We use essentially the same strategy herein by motivating this approach from a different perspective. Numerical problems are surpassed by incorporating the balance of angular momentum delivering the necessary smoothness for the pressure. Conventionally, the balance of angular momentum is neglected since it is already fulfilled locally by the balance of linear momentum (for the case of non-polar fluids). Furthermore, we discuss the integration by parts and suggest another approach than usually seen in the literature. We explain in detail how to generate the weak form. Additionally, we emphasize that the weak form can be extended to fluid structure interaction or multiphysics problems very easily. We use open-source packages developed under the FEniCS project\footnote{The FEniCS computing platform, https://fenicsproject.org/ } and solve some academic examples in order to present the accuracy, local monotonous convergence, and robustness of the proposed methodology. All codes are made public on the web site in \cite{compreal} to be used under the GNU Public license as in \cite{gnupublic} for promoting an efficient scientific exchange as well as further studies.   

\section{Variational formulation}

Consider the following general balance equation:
\begeq \label{gen.balance}
\Dt{ \Bigg( \int_\Om \psi \d\v \Bigg) } = \int_\Om z \d\v + \int_{\p\Om} (f+\phi) \d a \ , 
\eqend
where the rate of the variable $\psi$ is balanced with the supply term $z$ acting volumetrically and with the flux terms---convective $f$ and non-convective $\phi$---applying on the surface $\p\Om$ of a domain (control volume) $\Om$. By using Table~\ref{tab:supplyandfluxes}, we can obtain the balance equations of mass, linear momentum, and angular momentum. 
\begin{table}[!hbt]
\centering
\caption{Volume densities, their supply and flux terms in the balance equations.}
\renewcommand{\arraystretch}{1.3} 
\begin{tabular}{c|c|c|c}
\toprule
$\psi$ &
$z$ &
$f$ & 
$\phi$ \\
\midrule
$\rho$ & $0$ & $n_i (w_i-v_i) \rho $ & $0$ \\
$\rho v_j$ & $\rho g_j$ & $n_i (w_i-v_i)\rho v_j$ & $\sigma_{ij}$ \\
$\rho (s_j + \epsilon_{jkl} x_k v_l)$ & $\rho (\ell_j + \epsilon_{jkl} x_k g_l)$ & $n_i(w_i-v_i)\rho (s_j + \epsilon_{jkl} x_k v_l)$  & $m_{ij} + \epsilon_{jkl} x_k \sigma_{il}$\\
\bottomrule
\end{tabular}
\label{tab:supplyandfluxes}
\end{table}
The domain may have its own velocity, $\Dt{x_i} =w_i$, independent on the velocity of the fluid particle, $v_i$. For a discussion of the balance equations in a control volume with the domain velocity, we refer to \cite{MuellerI, MuellerII}. This domain velocity can be chosen arbitrarily without affecting the underlying physics. Herein we fix the domain by setting $w_i=0$. We assume that initially the fluid rests, $v_i(\t x, t=0)=0$, and it is a homogeneous material, $\rho(\t x, t=0)=\text{const}$. Moreover, we assume that the flow is incompressible, i.e., the mass density remains constant in time, $\p\rho/\p t=0$. After utilizing the \textsc{Gauss--Ostrogradskiy} theorem, we obtain
\begeq \label{governing1}
\int_{\Om} \rho \pd{ v_i }{x_i} \d\v = 0 \ , \\
\int_\Om \bigg( \rho \pd{v_j}{t} - \rho g_j + \rho \pd{v_i v_j}{x_i} - \pd{\sigma_{ij}}{x_i} \bigg) \d\v = 0 \ , \\
\int_\Om \bigg( \rho \pd{s_j}{t} + \rho \epsilon_{jkl} x_k \pd{v_l}{t} - \rho \ell_j - \rho \epsilon_{jki} x_k g_i 
- \pd{}{x_i} \big( -v_i \rho (s_j + \epsilon_{jkl} x_k v_l) + m_{ij} + \epsilon_{jkl} x_k \sigma_{il} \big) \bigg) \d\v =0 \ ,
\eqend
where the spin density, $s_i$, its flux term (couple stress), $m_{ij}$, and its supply term, $\ell_i$, they all vanish for a non-polar medium like water (furnishing a symmetric stress). Then the angular momentum is identical to the moment of (linear) momentum,
\begeq \label{governing2}
\int_\Om \epsilon_{jkl} x_k \bigg( \rho  \pd{v_l}{t} 
- \rho  g_l 
+ \rho  \pd{v_i v_l}{x_i}
- \pd{\sigma_{il}}{x_i}
\bigg) \d\v = 0
\ , \\
\int_\Om \bigg( \rho  \pd{v_l}{t} 
- \rho  g_l 
+ \rho  \pd{v_i v_l}{x_i}
- \pd{\sigma_{il}}{x_i}
\bigg) \d\v = 0
\ ,
\eqend 
hence, in analytical mechanics, we may neglect it. However, we observe this term as being important for resolving numerical challenges. 

We multiply the latter equations by an arbitrary test function with the same rank of the integrand, i.e., Eq.\,\eqref{governing1}$_1$ by a scalar, Eq.\,\eqref{governing1}$_2$ by a vector, and Eq.\,\eqref{governing2}$_2$ by a vector. Having an arbitrary test function ensures that the global condition holds locally as well. According to the \textsc{Galerkin} approach, we will choose the test functions from the same space as the unknowns,  $v_i$ and $p$. Therefore, it is natural to use $\del p$ and $\del v_i$ as possible test functions. As we need another vector as well, we may choose $\del v_i$ or even construct a vector by using gradient of $\del p$. By utilizing the latter, we will circumvent several numerical problems in the implementation, which is one of the key contributions of this work providing stability and robustness to the computational method. A possible justification of this observation relies on the restriction about the gradient of pressure by weighing Eq.\,\eqref{governing2}$_2$ by $\p\del p / \p x_l$, which is indeed the missing condition for the necessary numerical smoothness for pressure. We may see this condition related to the LBB condition; however, we enforce it by using an additional integral form instead of changing the order of shape functions. An analogous term multiplied by the mesh size is added in PSPG for the sake of a pressure stabilization. Herein we use it in equal manner for every finite element independent of their size. Therefore, this formulation is not called a stabilization since the same weak form is evaluated in each node with no dependence on the mesh size.

We begin utilizing the discrete representations of continuous fields; however, we omit a clear distinction in the notation since we never use continuous and discrete functions together. We emphasize that the choice of the scalar test function is critical and we suggest to use
\begeq \label{integral.forms}
\int_{\Om} \rho \pd{ v_i }{x_i} \del p \d\v = 0 \ , \\
\int_\Om \bigg( \rho \pd{v_j}{t} - \rho g_j + \rho \pd{v_i v_j}{x_i} - \pd{\sigma_{ij}}{x_i} \bigg) \del v_j \d\v = 0 \ , \\
\int_\Om \bigg( \rho  \pd{v_j}{t} 
- \rho  g_j 
+ \rho  \pd{v_i v_j}{x_i}
- \pd{\sigma_{ij}}{x_i}
\bigg) \pd{\del p}{x_j} \d\v = 0
\ .
\eqend
We refrain ourselves from inserting the mass balance, $\p v_i/\p x_i=0$, into the latter formulation. Since this condition is tested by $\del p$, we cannot expect that it is fulfilled for the velocity distribution, thus, we choose to enforce it by testing with $\del v_j$ in Eq.\,\eqref{integral.forms}$_2$. We emphasize that this particular point is overseen in many implementations leading to numerical errors in cases where the incompressibility condition becomes critical.

We solve the transient integral forms in discrete time slices with a time step $\Delta t$, this discretization in time is established by using \textsc{Euler} backwards method resulting in
\begeq\label{rate.def}
\pd{v_i}{t} = \frac{v_i-v_i^0}{\Delta t} \ ,
\eqend
where $v_i^0$ indicates the value from the last time step. This implicit method is stable (for real valued problems) and easy to implement. We stress that the implementation is an implicit method since we evaluate the derivative at the current time. Simply by using a \textsc{Taylor} series around the current time,
\begeq
v_i(t-\Delta t) = v_i(t) - \Delta t \pd{v_i}{t} + O(\Delta t^2) \ ,
\eqend  
and truncating after the linear term in $\Delta t$, we immediately obtain Eq.\,\eqref{rate.def}. As we neglect quadratic time steps, we have to choose small time steps in the simulation in order to increase the accuracy of the time discretization. 

\section{Generating the weak form}

Integral forms in Eq.\,\eqref{integral.forms} will be rewritten in the same unit such that we can sum them up. First, we divide Eq.\,\eqref{integral.forms}$_1$ by the mass density $\rho$ and bring it to the unit of power. Second, we undertake no changes in Eq.\,\eqref{integral.forms}$_2$ as it is already in the unit of power. Third, we divide Eq.\,\eqref{integral.forms}$_3$ by the mass density and multiply it by the time step $\Delta t$ and bring it also to the unit of power. The choice of the unit, herein the unit of power, is arbitrary. Often the formulation is employed in a dimensionless form generating the well-known \textsc{Reynolds} number. 

We rewrite the constitutive equation, $\sigma_{ij} = -p\delta_{ij} + \tau_{ij}$, by combining all terms with $d_{ij}$ into $\tau_{ij}$. The symmetric part of the velocity gradient, $d_{ij}$, thus the term $\tau_{ij}$  already include velocity's first derivative in space. As a consequence of the first derivative, we need to have a representation of the velocity function, which is at least $C^1$ continuous. In the integral forms, we observe another space derivative in $\tau_{ij}$ leading to the restriction that a velocity approximation belonging to the class $C^2$ has to be implemented. This condition will be weakened by integrating by parts. We emphasize that the integration by parts is applied \textit{only} on the terms having (at least) second derivative of the unknown in order to ``shift'' one differentiation to the test function. This strategy is not conventional. Often, integration by parts is used to all flux terms. We observe numerical problems by using an integration by parts employed to all flux terms. We suggest to integrate by parts only if the term consists of a second derivative of the unknown. This approach is another key contribution of the work. After bringing to the same unit and integrating by parts where necessary, by using the usual comma notation for space derivative $()_{,i}=\p()/\p x_i$, we obtain the following weak forms:
\begeq
F_1 = \int_\Om v_{i,i} \del p \d\v 
\ , \\
F_2 = \int_\Om \bigg( 
\rho \frac{v_j-v_j^0}{\Delta t} \del v_j 
- \rho g_j \del v_j 
+ \rho (v_i v_j)_{,i} \del v_j
+ p_{,i} \del v_i
+ \tau_{ij} \del v_{j,i}  \bigg) \d\v 
- \int_{\p\Om} n_i \tau_{ij} \del v_j \d a
 \ , \\
F_3 = \int_\Om \Big( 
 (v_j-v_j^0) 
 - \Delta t g_j 
 + \Delta t (v_i v_j)_{,i} 
 - \frac{\Delta t}{\rho} (-p \delta_{ij} + \tau_{ij})_{,i}  \Big) \del p_{,j} \d\v
\ .
\eqend
Summing them reads the weak form to be solved,
\begeq
\text{F}\big|_\Om = F_1+F_2+F_3 \ ,
\eqend
for one finite element. It becomes zero by inserting the correct pressure and velocity distribution. A control volume is decomposed into several elements, by assembling the weak forms of each finite element, $\text{Form} = \sum_{e} F\big|_\Om$, the weak form for the whole control volume is acquired. This assembling generate on the element boundaries the following term:
\begeq
\jl n_i \tau_{ij} \jr 
= \jl n_i (\sigma_{ij}+p\delta_{ij}) \jr \ , 
\eqend
with jump brackets $\jl (\cdot) \jr$ indicating the difference between the values of the quantity computed in adjacent elements. We use continuous form elements for pressure and velocity such that $\jl n_i p \jr = 0$, moreover, we enforce $\jl n_i \sigma_{ij} \jr = 0$ relying on the balance of linear momentum on singular surfaces. Therefore, the integral term along the element boundaries vanish within the control volume. On the boundaries of the control volume either velocity or pressure is given. For the parts, where velocity is given, we choose $\del v_i=0$ such that the boundary term vanishes. For the boundary parts, for example where fluid enters or leaves the domain with a prescribed pressure, $p$, against the plane outward direction, $t_i = -p n_i$, we obtain $n_i \tau_{ij} = n_i (\sigma_{ij} +  p \delta_{ij}) = t_j + n_j p = 0$. Therefore, all boundary terms vanish in the weak form for an incompressible flow.

\section{Algorithm and computation}

Continuous finite elements are used for all simulations. In three-dimensional space (3D) we use tetrahedrons as elements and in two-dimensional space (2D) we use triangles as elements---both with linear form functions, i.e., form functions of degree $n=1$. The primitive variables are pressure and velocity; they are represented with corresponding nodal values interpolated using the form functions. Concretely, 4 primitive variables $\t P = \{p, v_1, v_2, v_3\}$ in three-dimensional space belong to
\begeq
\mathcal V = \big\{ \t P \in [\mathcal H^n(\Om)]^4 : \t P|_{\p\Om}=\text{given} \big\} \ ,
\eqend
where $[\mathcal H^n]^4$ is a 4-dimensional \textsc{Hilbert} space of class $C^n$ as defined in \cite{hilbert} with additional differentiability properties such that it is called a \textsc{Sobolev} space. The test functions, $\del \t P=\{\del p, \del v_1, \del v_2, \del v_3\}$, stem from the same space
\begeq
\hat{\mathcal V} = \big\{ \del \t P \in [\mathcal H^n(\Om)]^4 : \del \t P|_{\p\Om}=\text{given} \big\} \ ,
\eqend
which is the \textsc{Galerkin} approach. We use $n=1$ for all simulations, in other words, we use the same linear form functions for pressure as well as for velocity. This choice is risky since we fail to fulfill the aforementioned LBB conditions. Despite this fact, at least for the demonstrated applications, no spurious oscillations occur owing to the additional governing equation restricting the pressure gradient as well as the careful choice of terms for integrating by parts. 

The weak form is nonlinear and coupled such that we need to linearize and solve it monolithically. For the linearization, we follow the ideas in \cite[Part I, Sect.\,2.2.3]{logg2012} and perform an abstract linearization using \textsc{Newton}'s method at the partial differential level. The functional Form$=F(\t P,\del\t P)$ is an integral of the function depending on the primitive variables $\t P$ and their variations (test functions) $\del \t P$. We know the correct values of $\t P$ at $t=0$. The weak form is initially zero---we obtained it by subtracting left-hand sides from right-hand sides in the balance equations. We search for $\t P(t+\D t)$ at the next time step, $\D t+ t$, by using the known values $\t P(t)$. This algorithm holds for every time steps, since we compute subsequently in time. We describe the algorithm as follows:
\begin{equation}
\begin{aligned}
\text{given: } & \t P(t) \text{  for  } \t x \ , \\
\text{find: } & \t P(t+ \D t) \text{  at  } \t x\ , \\ 
\text{satisfying: } & F(\t P(t+\D t), \del \t P)=0 \ .
\end{aligned}
\end{equation}
Now, by rewriting the unknowns $\t P(t+\D t)$ in terms of the known values
\begeq
\t P(t+\D t) = \t P(t) + \D\t  P(t) \ ,
\eqend 
we redefine the objective to searching for $\D \t P(t)$ instead of $\t P(t+\D t)$. If $\D t$ is chosen sufficiently small, then the solution is near to the known solution such that $\D\t P(t)$ is small. This condition leads to a \textsc{Taylor} expansion around the known values, $\t P(t)$, up to the (polynomial) order one
\begeq
F(\t P+\D\t P, \del\t P) = F(\t P, \del\t P) + \grad_{\t P} F(\t P,\del\t P)  \cdot \D\t P \ ,
\eqend
where we omit the time argument for the sake of clarity in notation. The expansion is linear in $\D\t P$, hence we need to construct a linear in $\D\t P$ differentiation operator, $\grad_{\t P}$, which is established by the so-called \textsc{Gateaux} derivative:
\begeq
\grad_{\t P} F(\t P,\del\t P) \cdot \D\t P
= \lim_{\epsilon\rightarrow 0} \dd{ }{\epsilon} F(\t P+\epsilon\D\t P,\del\t P) \ ,
\eqend 
where $\epsilon$ is an arbitrary parameter. Since we first differentiate in $\epsilon$ and then set the parameter $\epsilon$ equal to zero, only terms of order one in $\D\t P$ remain in the solution. By introducing the so-called \textsc{Jacobi}an:
\begeq
\t J(\t P,\del\t P) = \grad_{\t P} F(\t P,\del\t P) \ ,
\eqend
we rewrite the algorithm,
\begin{equation}
\begin{aligned}
\text{given: } & \t P \text{  for  } \t x \ , \\
\text{find: } & \D \t P \text{  at  } \t x\ , \\ 
\text{satisfying: } & F(\t P, \del \t P) + \t J(\t P, \del \t P) \cdot \D \t P =0 \ . 
\end{aligned}
\end{equation}
The last line is a linear function in $\D\t P$ such that we can solve the equation by obtaining $\D\t P$ and update the solution in an iterative manner,
\begeq
\t P := \t P+\D\t P \ ,
\eqend
where ``$:=$'' is an assign operator in computational algebra. Here is the ultimate algorithm:
\begin{equation}
\begin{aligned}
\text{while  } &|\Delta \t P |>\text{TOL.} \\
 & \text{ solve } \Delta \t P \text{, where }  F(\t P, \del \t P) + \t J(\t P, \del \t P) \cdot \D \t P =0 \\
 &\t P := \t P + \Delta \t P 
\end{aligned}
\end{equation}
The term $\t J \cdot \D \t P$ is computed automatically by means of symbolic differentiation implemented under the name SyFi within the FEniCS project, see \cite{AlnaesMardal2010a}, \cite{AlnaesMardal2012b}. This automatic linearization procedure allows us to use any nonlinear constitutive equation in the code. Herein we use a linear constitutive equation in order to achieve a comparison with closed-form solutions. For higher \textsc{Reynolds} numbers, the procedure may lead to a slow or even non-convergence. This numerical problem is caused by the linearization itself as discussed in \cite[Sect. 7.2]{elman2005}. As we have expanded with a linear \textsc{Taylor} series, the initial guess of the \textsc{Newton--Raphson} algorithm affects the convergence greatly. A better option is to use the \textsc{Picard} (fix point) iteration with the same weak formulation as introduced herein.

The geometry is constructed in Salome\footnote{Salome, the open-source integration platform for numerical simulation, http://www.salome-platform.org } by using NetGen algorithms\footnote{Netgen Mesh Generator, https://sourceforge.net/projects/netgen-mesher/ } for the triangulation. Then the mesh is transformed as explained in \cite[Appendix A.3]{027} and implemented in a Python code using packages developed by the FEniCS project, which is wrapped in C++ and solved in a Linux machine running Ubuntu.\footnote{Ubuntu, open source software operating system, https://www.ubuntu.com/ } 

\section{Comparative analysis}

The suggested weak form is implemented and solved for various problems. First, we examine the accuracy and convergence behavior by comparing to semi-analytical closed-form solutions for simple geometries, we call them analytical solutions. They are all well-known and can be found in different textbooks, for example see \cite{savas_me260}. This analysis is of importance to present the local monotonous convergence that is the important and prominent feature of the FEM. For a flow problem, we increase the accuracy at every point by decreasing the mesh size. Second, we present benchmarking problems in 2D and 3D in order to verify the robustness of the method.

\subsection{Steady state \textsc{Hagen--Poiseuille} flow}

Consider a laminar flow in an infinite pipe as a result of the given pressure difference. This configuration is called the \textsc{Hagen--Poiseuille} flow and it has a steady-state solution obtained in cylindrical coordinates, $r$, $\theta$, $z$, under the assumption that the flow is only along the pipe. The pipe is oriented along $z$, which is set as the axis of the pipe. No-slip condition, $v_i=0$, is applied on the outer walls, $r=a$. The steady-state solution reads
\begeq
v_i = \begin{pmatrix} 0 \\ 0 \\ \displaystyle -\dd{p}{z} \frac{a^2}{4\mu} \Big( 1-\frac{r^2}{a^2} \Big) \end{pmatrix} \ ,
\eqend 
for an incompressible flow of a linear viscous fluid like water of viscosity $\mu$. We use this solution for an analysis of the convergence and accuracy. A three-dimensional pipe of length $\ell$ is constructed and on the inlet and outlet, the pressure is given as \textsc{Dirichlet} boundary conditions, $p(z=0)=p_\text{in}$ and $p(z=\ell)=p_\text{out}$. The flow is driven by the pressure difference such that the gravity is neglected. Moreover, we are interested in the steady-state where the inertial term vanishes. In order to mimic the infinite pipe, we set the radial and circumferential velocities zero on the inlet and outlet. FEM computation is realized in Cartesian coordinates, so we basically allow $v_z$ on inflow plane, $x=0$, and outflow plane, $x=\ell$, by setting $v_x=v_y=0$ as \textsc{Dirichlet} conditions.

The pressure distribution is expected to be linear along $z$, hence, for the analytical solution $\d p/\d z=(p_\text{out}-p_\text{in})/\ell$. For a better comparison we use the diameter $D$ as the characteristic length and half of the maximum velocity, $v_z^\text{M}=v_z(r=0)$, as the characteristic velocity for calculating the \textsc{Reynolds} number:
\begeq
Re = \frac{v_z^\text{M} D \rho}{2 \mu} \ .
\eqend
For a small pipe of an inch long and a quarter inch wide, we construct the mesh by using a global element length $h$. We use SI units such that $\ell=25.4$\,mm and $D=6.35$\,mm and water as the fluid with
\begeq \label{water.parameters}
\rho = 998.2\cdot 10^{-6}\,\text{g/mm$^3$} \comma
\mu = 1001.6\cdot 10^{-6}\,\text{Pa\,s} \comma
\lambda = 0.6\,\text{Pa\,s} \ .
\eqend  
Especially the choice of $\lambda$ is of importance since this parameter is not measured directly. Although $\lambda$ is a physical and measurable quantity, for the conventional pressure the incompressibility makes the measurement of $\lambda$ very challenging. In the constitutive equation in Eq.\,\eqref{constitutive}, the term $\lambda d_{kk}$ vanishes for the correct velocity solution. Therefore, for the numerical sense, $\lambda$ has to be great enough that $d_{kk}$ is enforced to vanish. Hence, we choose $\lambda$ multiple times greater than $\mu$, in reality (for compressible flows) $\lambda$ and $\mu$ are independent parameters. Interestingly, we have observed that choosing $\lambda$ greater than suggested value slows down the convergence. A remedy to this relies on the aforementioned fixed point iteration; but we use herein the \textsc{Newton--Raphson} iteration. By doubling $\lambda$, the same accuracy of the solution is obtained with more degrees of freedom (DOF). Its value does not change the convergence behavior, as long as it is great enough. In order to determine the value of $\lambda$, we simply decreased until the maximum $Re$ was achieved. Less than the used value leads to numerical problems in the \textsc{Newton--Raphson} iterations. 

\begin{figure}
\centering
\subfloat[$Re = 313$]{\includegraphics[width=0.7\textwidth]{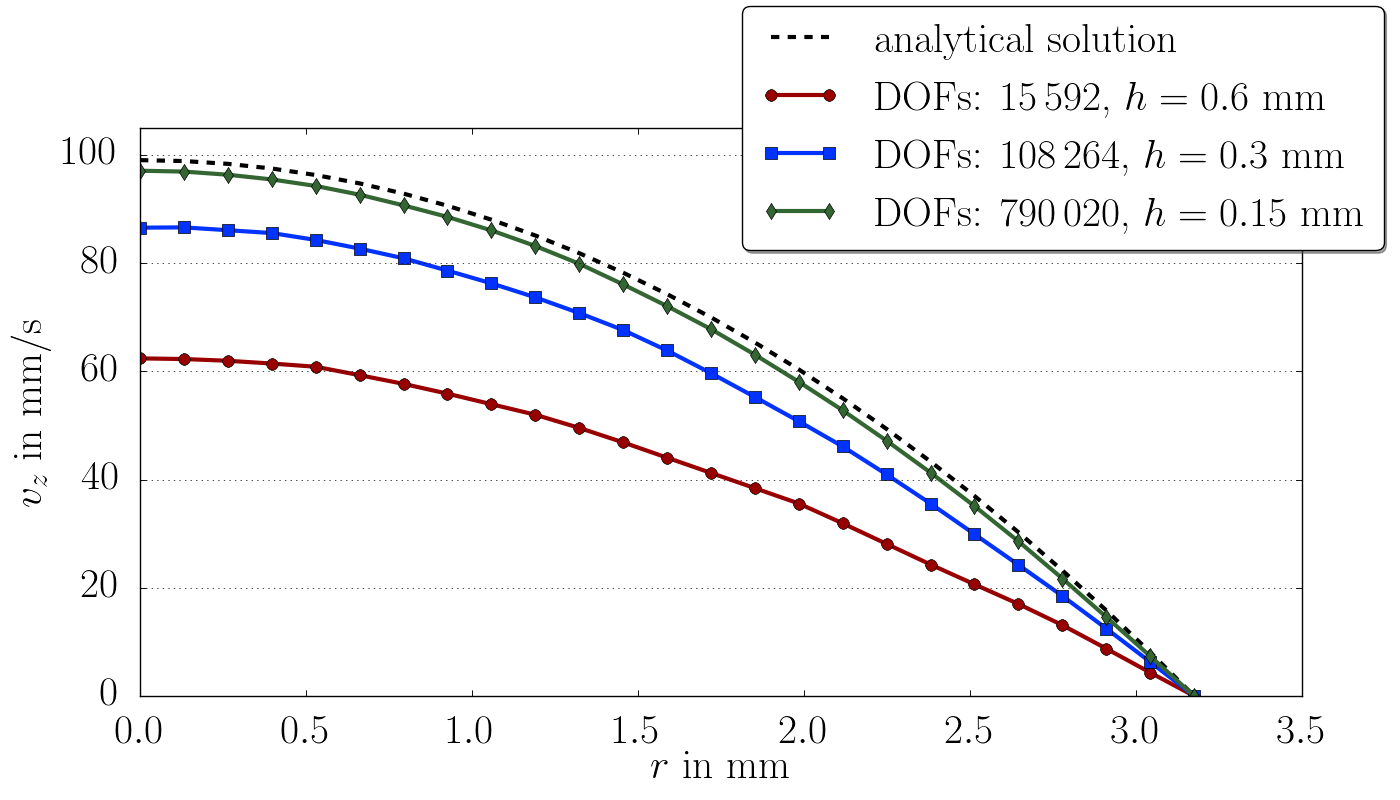}} \\
\subfloat[$Re = 940$]{\includegraphics[width=0.7\textwidth]{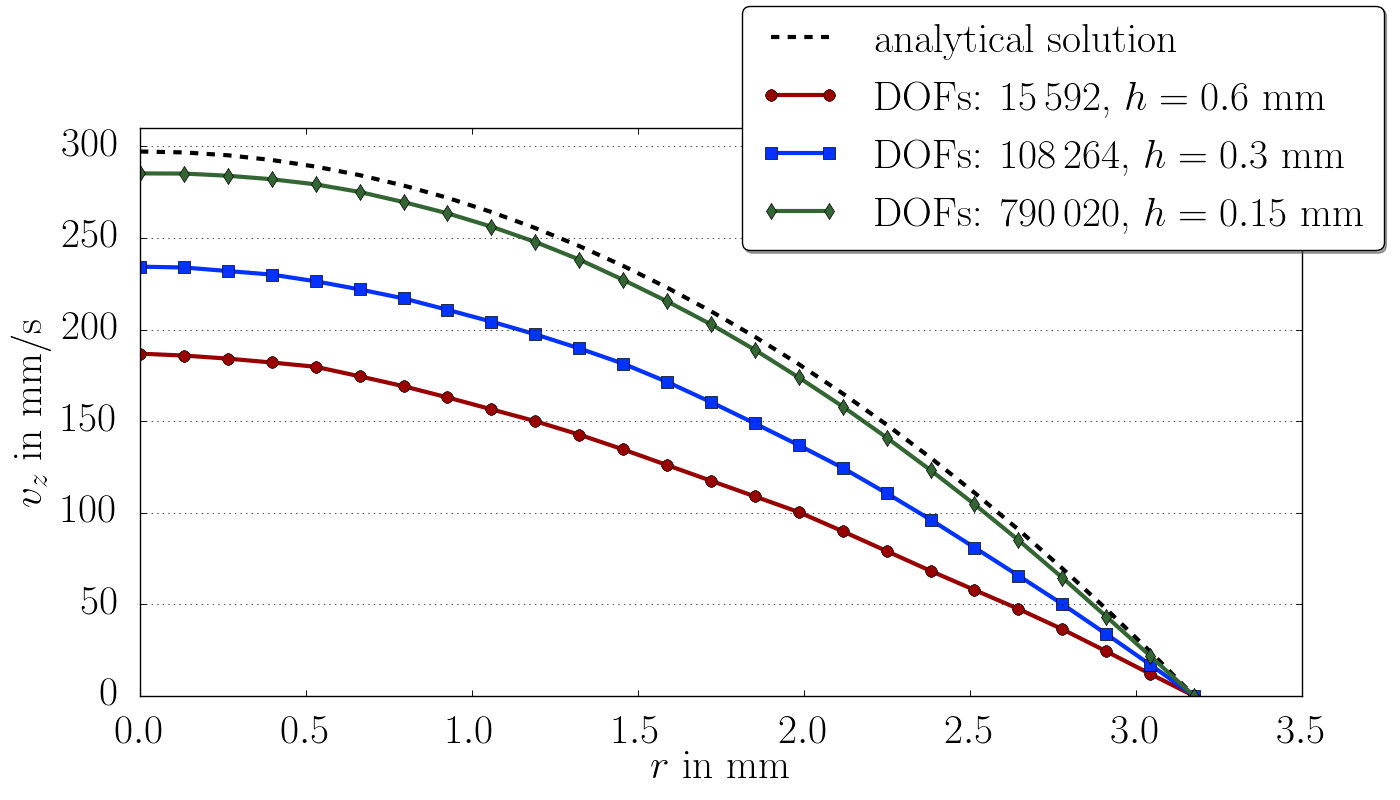}}
 \caption{3D computation of steady state solution and its comparison to the analytical solution in a pipe for two different \textsc{Reynolds} numbers.}
 \label{fig:steadystate} 
\end{figure}

By using a standard convergence analysis, we compare three different meshes starting with a global edge length of tetrahedrons, $h=0.6$\,mm, then reducing it by half. For every simulation, a new mesh is generated such that the number of nodes fail to increase exactly by $2^3$ times in 3D. We use an unstructured mesh and the mesh quality is nearly identical because of using the same algorithm on the relatively simple geometry. The expected monotonic convergence has been attained as seen in Fig.\,\ref{fig:steadystate} for 2 different \textsc{Reynolds} numbers.
Two important facts need to be underlined. First, the parabolic distribution of the velocity along the diameter is achieved even with a coarse mesh. Second, the relative error is 1.1\% at $r=0$ for a low \textsc{Reynolds} number and 4.1\% at $r=0$ for a high \textsc{Reynolds} number by using mumps direct solver.

\subsection{Starting \textsc{Hagen--Poiseuille} flow}

\begin{figure} 
\centering
\subfloat[$\Delta t=\{1, 0.5, 0.25\}$\,s]{\includegraphics[width=0.7\textwidth]{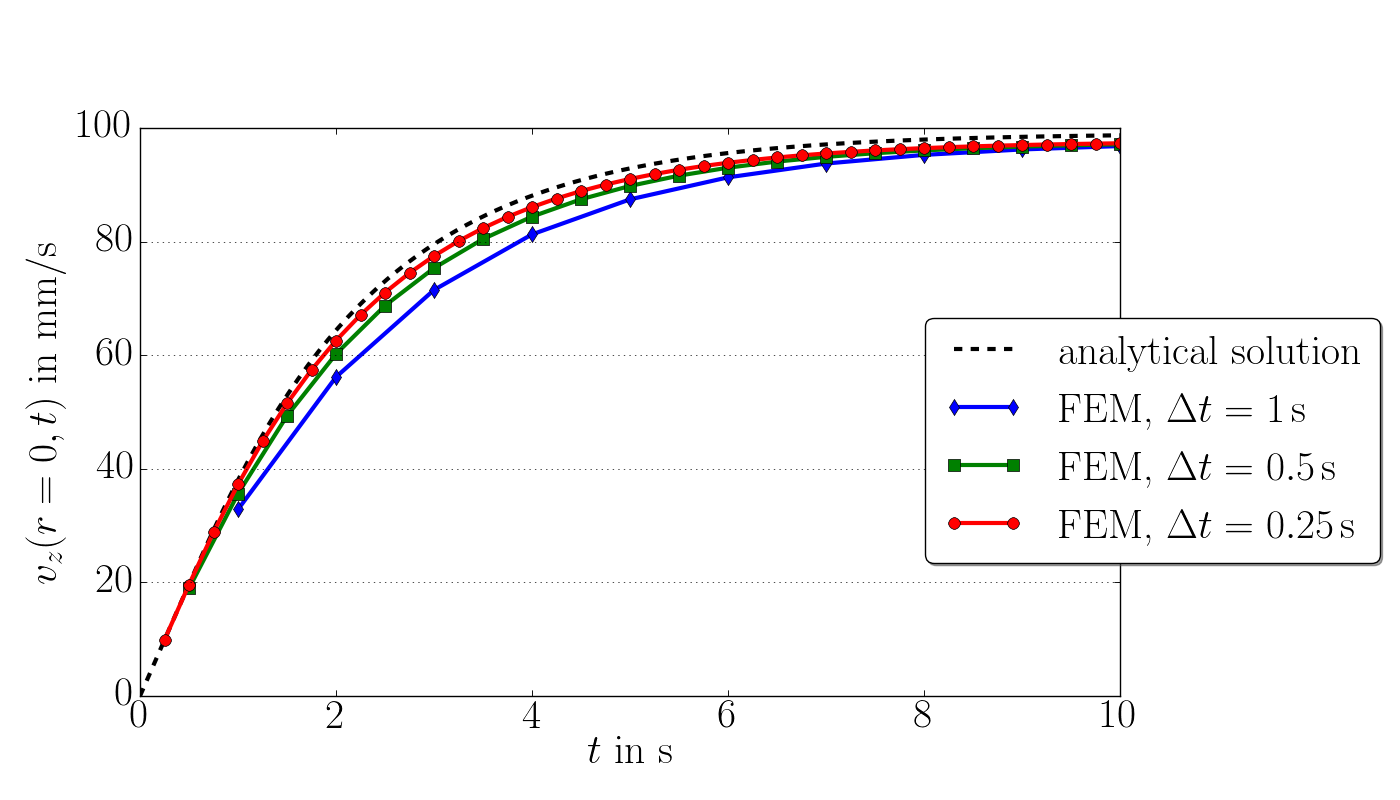}} \\
\subfloat[$\Delta t=0.25$\,s]{\includegraphics[width=0.7\textwidth]{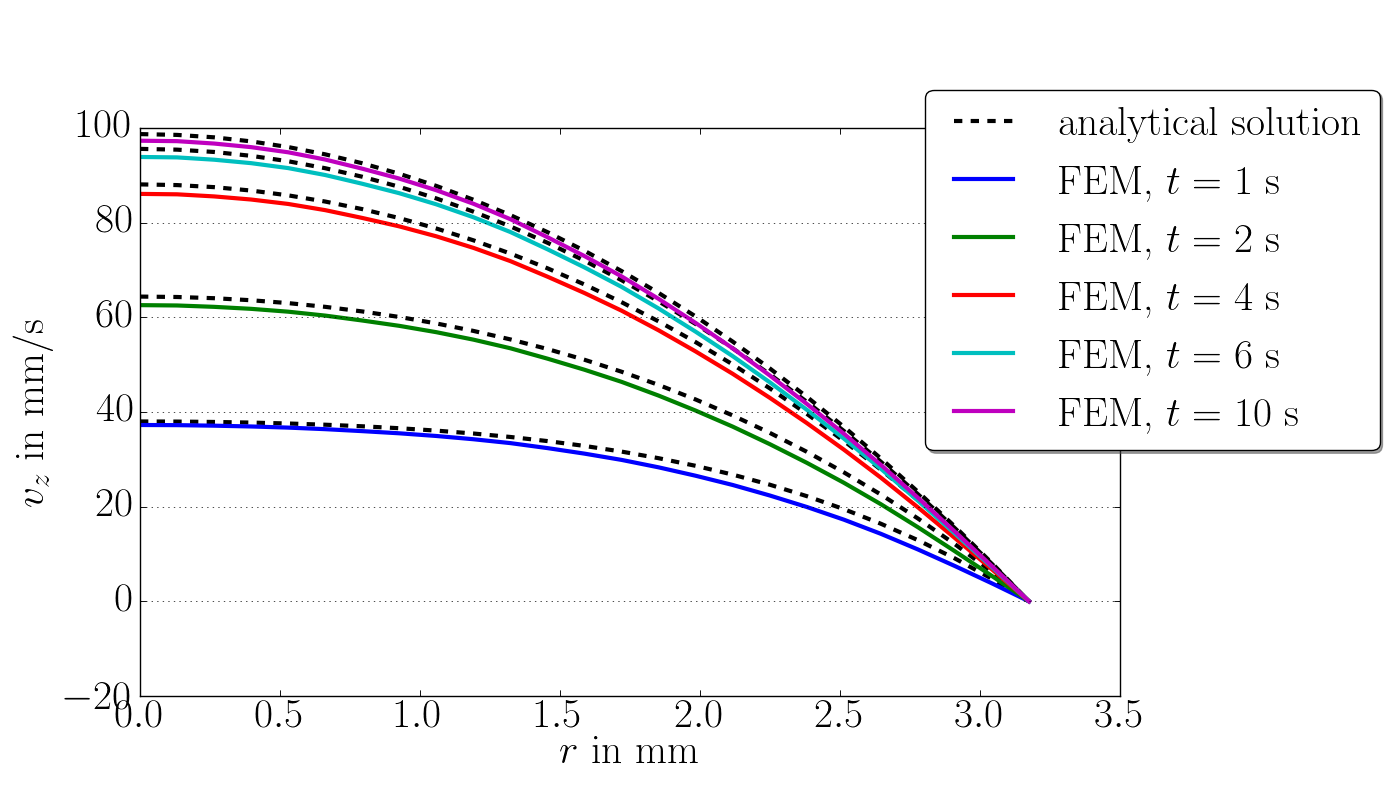}}
 \caption{3D computation of transient solution and its comparison to the analytical solution in a pipe for $Re=313$.}
 \label{fig:transient} 
\end{figure}

\begin{figure} 
\centering
\includegraphics[width=0.6\textwidth]{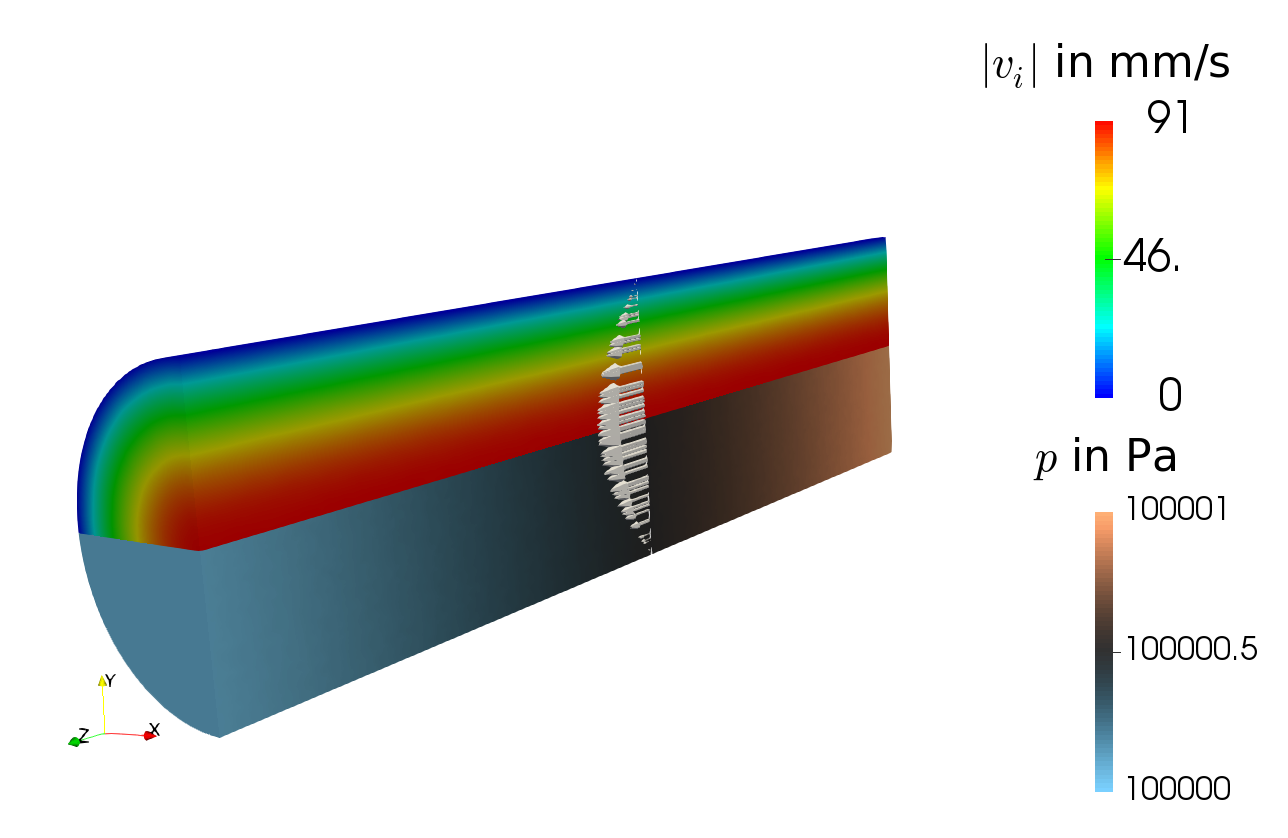}
 \caption{Velocity solution (direction as scaled arrows and magnitude as colors) and pressure solution (as colors) of the transient pipe flow at $t=5$\,s, shown on the half of the pipe (upper part: velocity and lower part: pressure) for $Re=313$.}
 \label{fig:velocity.pressure} 
\end{figure}

In order to test the accuracy in the transient simulation, we use the same configuration and solve it transiently in time. Since we have obtained the expected convergence in space discretization for the steady-state solution, we expect to have a monotonic convergence in time discretization, too. Therefore, we solve the same example with different time steps from $t=0$ to $t=10$\,s with the initially applied pressure difference. For this case there is a closed-form solution under the same assumptions as before,
\begeq
v_i = \begin{pmatrix} 0 \\ 0 \\ \displaystyle -\dd{p}{z} \frac{a^2}{4\mu} \bigg( 1-\frac{r^2}{a^2} - \sum_{n=1}^N 8 \frac{J_0(\Lambda_n\frac{r}{a})}{\Lambda_n^3 J_1(\Lambda_n) } \exp(-\Lambda_n^2 \tau) \bigg) \end{pmatrix} \ ,
\eqend 
with $\tau=t\mu/(\rho a^2)$ and \textsc{Bessel} functions (of the first kind) $J_0$ and $J_1$ with $\Lambda_n$ being the roots of $J_0$. We compute the (semi-)analytical solution by using SciPy packages for \textsc{Bessel} functions as well as its roots and choose $N=50$. By choosing the best mesh obtained from the convergence analysis in the steady-state case, namely the mesh with $h=0.15$\,mm, we compute the solution transiently in time by using different time steps. We present in Fig.\,\ref{fig:transient} the maximum value $v_z(r=0,t)$ over time for different time steps showing the expected convergence with decreasing time steps as well as the distribution at different time instants for the smallest time step.
In addition to the convergence in time, the relative error remains unchanged over time. We conclude that the suggested formalism is capable of simulating a simple, laminar, three-dimensional flow of a linear viscous fluid accurately under a monotonic loading. The velocity and pressure distributions show no artifacts or mesh dependency as presented in Fig.\,\ref{fig:velocity.pressure}.
We have used mumps direct solver for achieving the highest accuracy in the numerical solution. We emphasize that the pressure difference is applied instantaneously, which is numerically challenging. For transient loading scenarios, there are various assumptions used for obtaining a closed-form solution such that we omit to examine further cases. Based on the presented examples, we conclude that the approach delivers an accurate solution.

\subsection{Lid driven cavity}

\begin{figure} 
\centering
\includegraphics[width=0.7\textwidth]{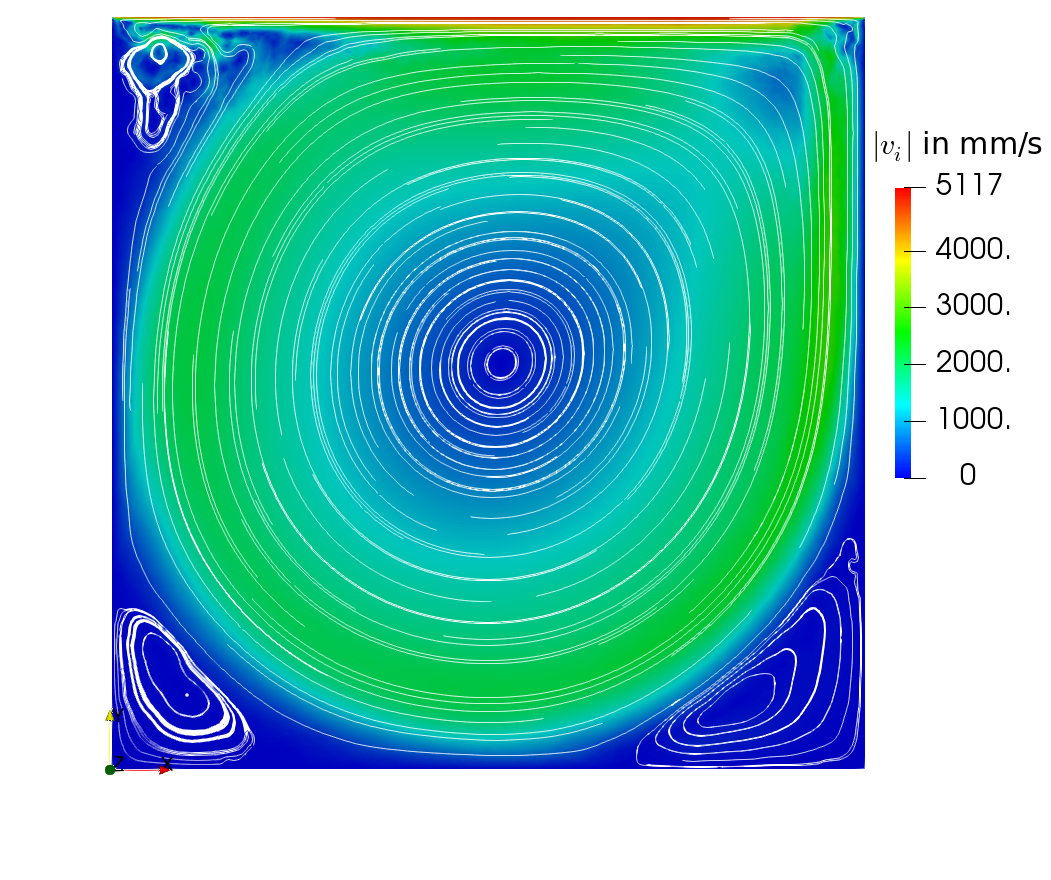}
 \caption{Velocity distribution and the corresponding streamline for $Re=5000$ simulated with parameters of water from Eq.\,\eqref{water.parameters}.}
 \label{fig:liddriven.streamlines} 
\end{figure}

\begin{figure} 
\centering
\includegraphics[width=0.7\textwidth]{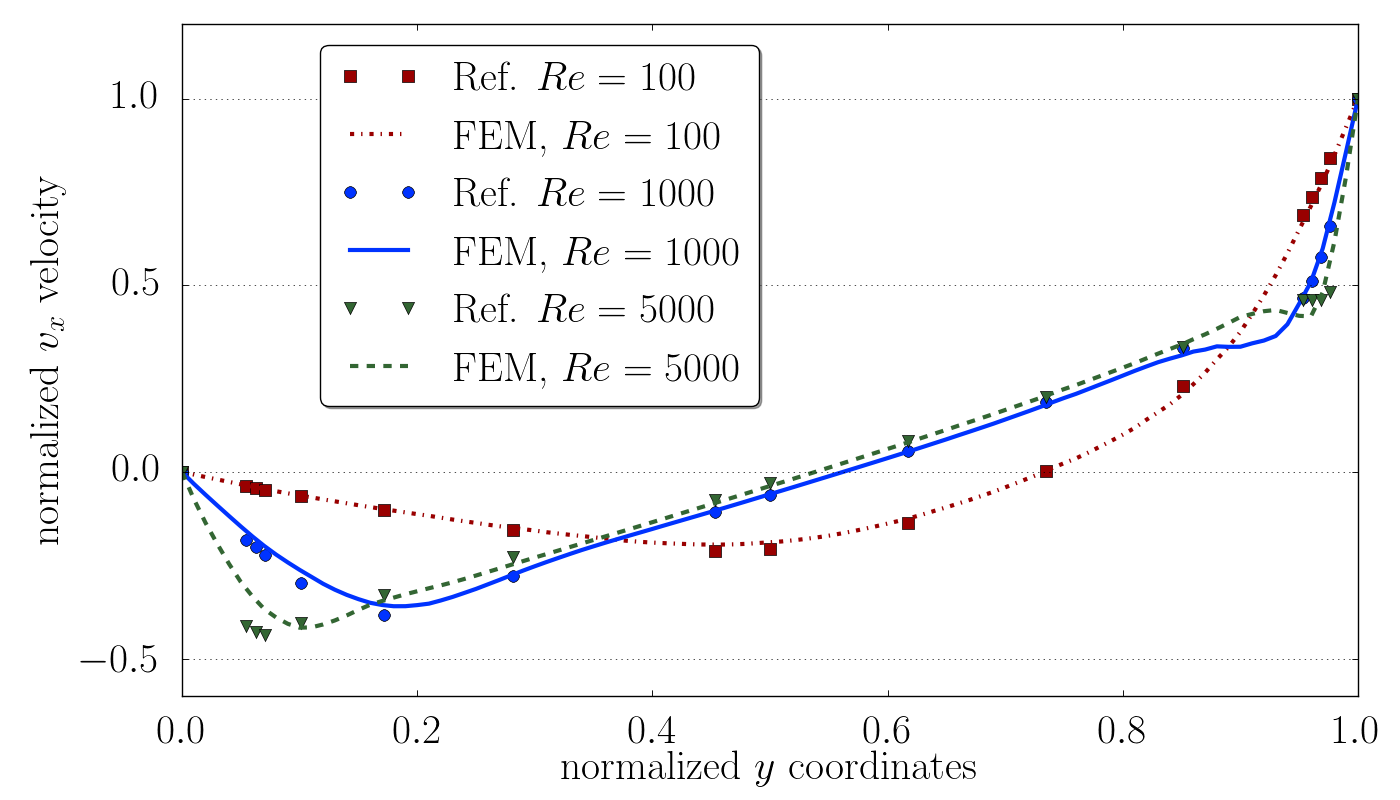}
\includegraphics[width=0.7\textwidth]{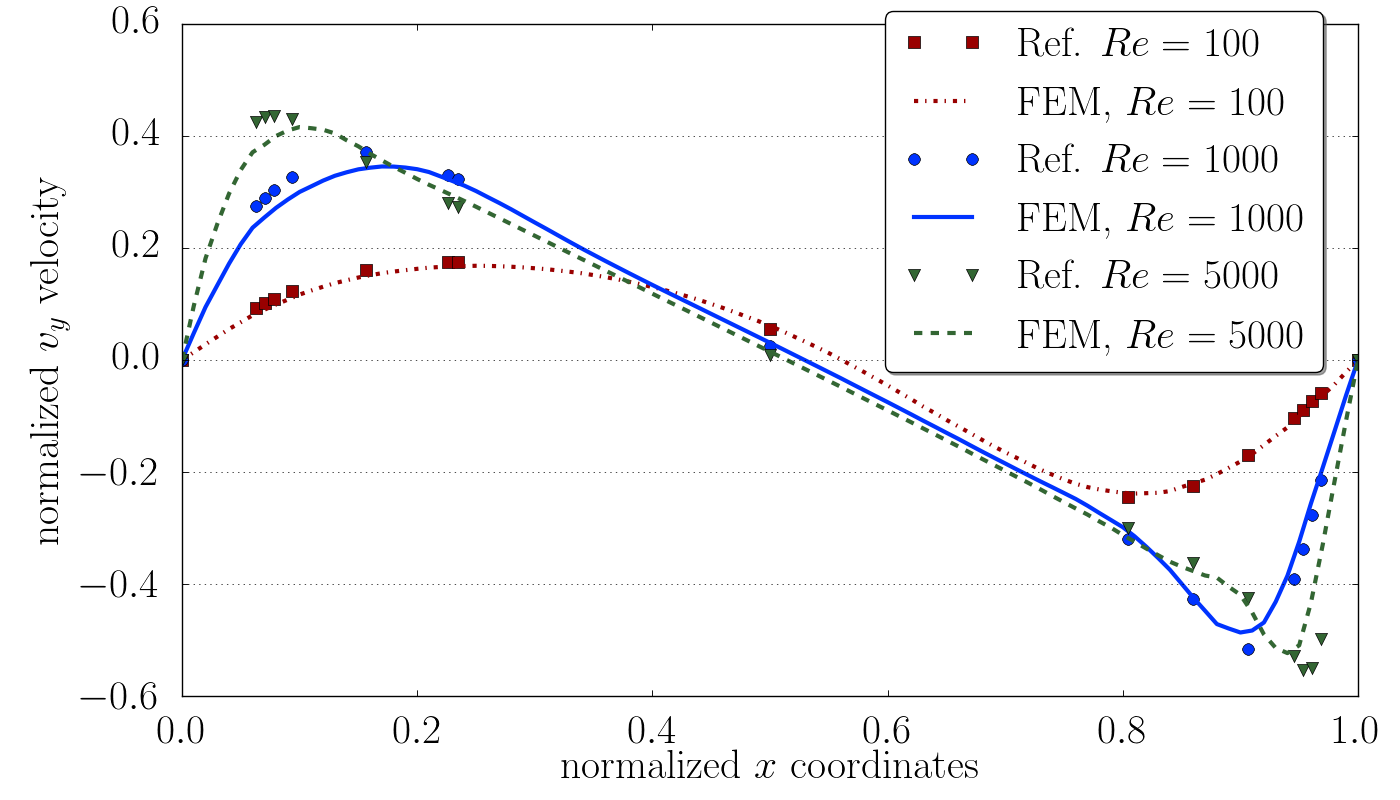}
 \caption{Velocity distribution in the lid driven cavity benchmark problem. Top: horizontal velocities along the vertical axis through the center are divided by the (given) lid velocity. Bottom: vertical velocities along horizontal axis through the center are normalized by the lid velocity. For three different \textsc{Reynolds} numbers, the results are compared to the reference solution (denoted as ``Ref.'') taken from \cite[Tables I and II]{ghia1982high}.}
 \label{fig:liddriven.compare} 
\end{figure}

Especially in higher \textsc{Reynolds} numbers, the stability of the numerical method becomes critical. Hence, we examine another mostly used benchmark problem by following \cite{ghia1982high}, where the numerical solution is obtained by using a different numerical solution strategy. We model a rectangle with a given shearing velocity on top, called lid driven cavity. As fluid we use water with parameters as in Eq.\,\eqref{water.parameters}; the solution is shown in normalized units for the sake of a direct comparison to \cite{ghia1982high} that we use as the reference solution. Without numerical problems, we managed to compute up to $Re=10\,000$ by a transient computation, where the shear velocity on the lid is slowly increased; so we may assume that the solution in each time step is tantamount to the steady state solution, for which the results are compiled in \cite[Tables I and II]{ghia1982high}. As an example we demonstrate in Fig.\,\ref{fig:liddriven.streamlines} the solution for $Re=5\,000$.
As is seen in Fig.\,\ref{fig:liddriven.streamlines}, the velocity distributions along horizontal and vertical axes may be used for comparing results adequately as presented in Fig.\,\ref{fig:liddriven.compare}.
Results from the reference solution and results obtained by the proposed method show no significant difference up to $Re=5\,000$. This good agreement fails to be the case in the higher \textsc{Reynolds} number, we skip an analysis about the strength and weak points of the method used for the reference solution. We emphasize that the proposed method is delivering results even for high \textsc{Reynolds} numbers. 

Typical numerical stability problems occurring in the finite element method are circumvented by using the aforementioned weak form as well as keeping the volume viscosity. For each application, the volume viscosity affects the numerical stability. We have chosen $\lambda = 10$\,Pa\,s. Considering the analysis before, we emphasize that results indeed show the same local and monotonous convergence properties. By choosing $\lambda$ bigger, we slow down the convergence such that a dense mesh is used for acquiring reliable results.

\subsection{\textsc{Karman} vortex street}

Flow past an obstacle is one of the heavily studied phenomena in the literature. Especially the instabilities in wakes behind an obstacle---realized as a bluff in the middle of the flow---has obtained much attention, we refer to \cite{williamson1996vortex} for a detailed historical review. By precisely setting up an arrangement as analyzed in \cite{karman1912mechanismus}, it is possible to obtain a \textit{vortex street} configuration depending on the force dragging and lifting the obstacle. For example, such a problem is solved in \cite[Section 3.4]{langtangen2016} by using the so-called incremental pressure correction scheme (IPCS). This splitting method is powerful for isothermal case, but difficult to apply for non-isothermal cases. In the approach presented herein, we solve pressure and velocity at once by using the same order of form functions. As suggested in \cite{schafer1996}, we implement a benchmarking problem for creating a laminar \textsc{Karman} vortex street. This benchmark problem is used to test a new method or code, it is accurately computed in \cite{john2004reference}, by using \textsc{Taylor--Hood} elements as quadratic form functions for velocity and linear functions for pressure, without stabilization, which we use as the reference solution. For a qualitative comparison with presented solutions in \cite[Fig. 2]{john2004reference}, at the same time instants, we visualize velocity distributions in Fig.\,\ref{fig:over.cyl.mesh2D} .
\begin{figure}
\centering
\includegraphics[width=0.85\textwidth]{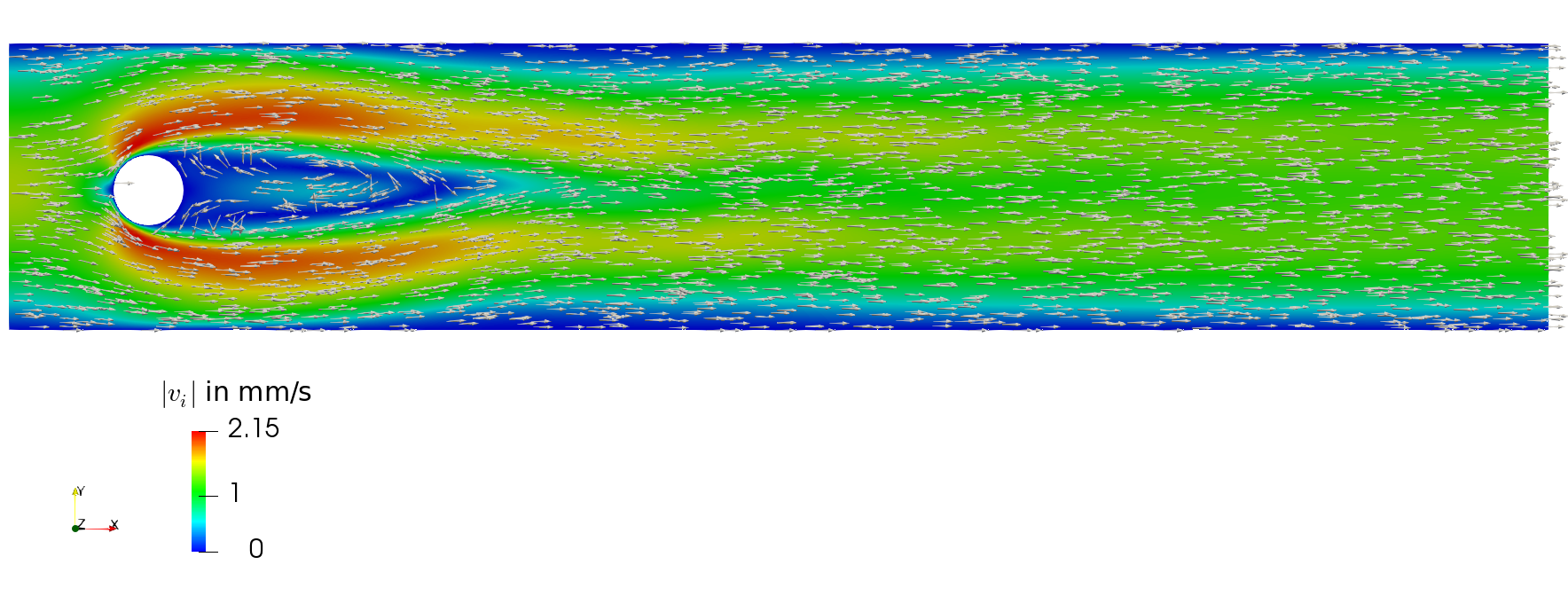}
\vspace{-7mm} \\
\includegraphics[width=0.85\textwidth]{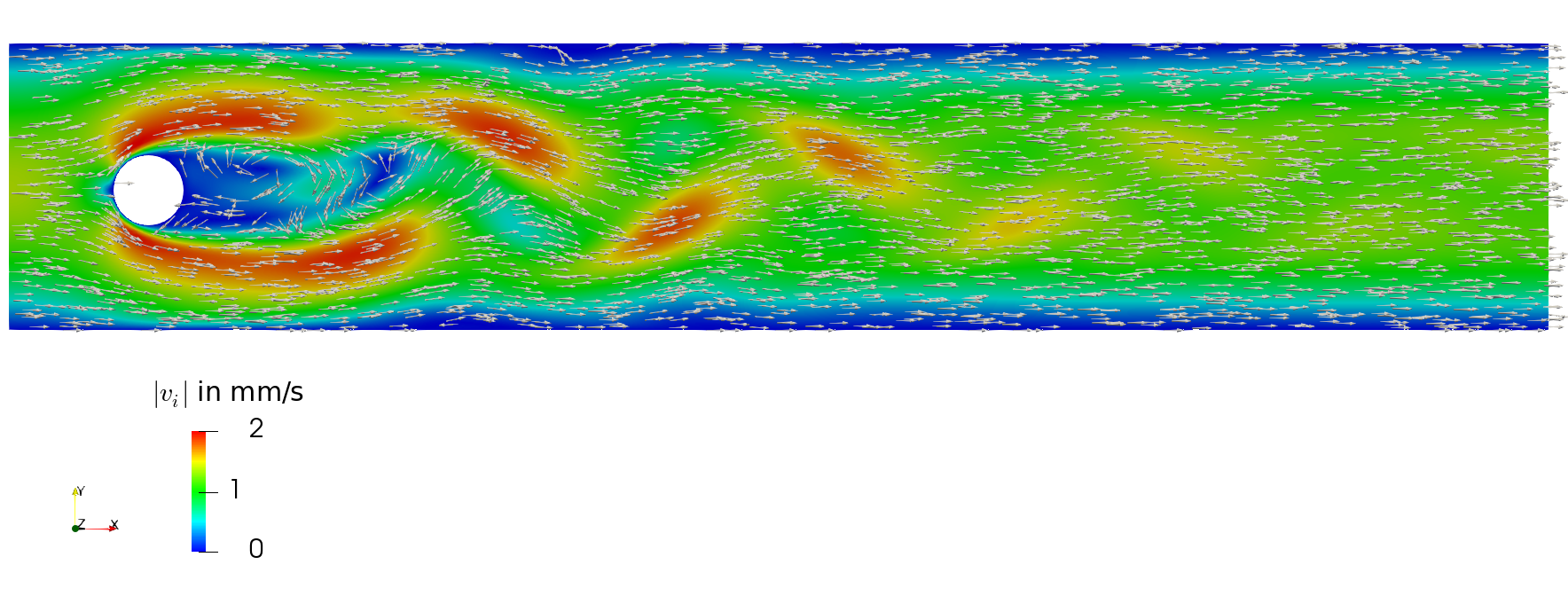}
\vspace{-7mm} \\
\includegraphics[width=0.85\textwidth]{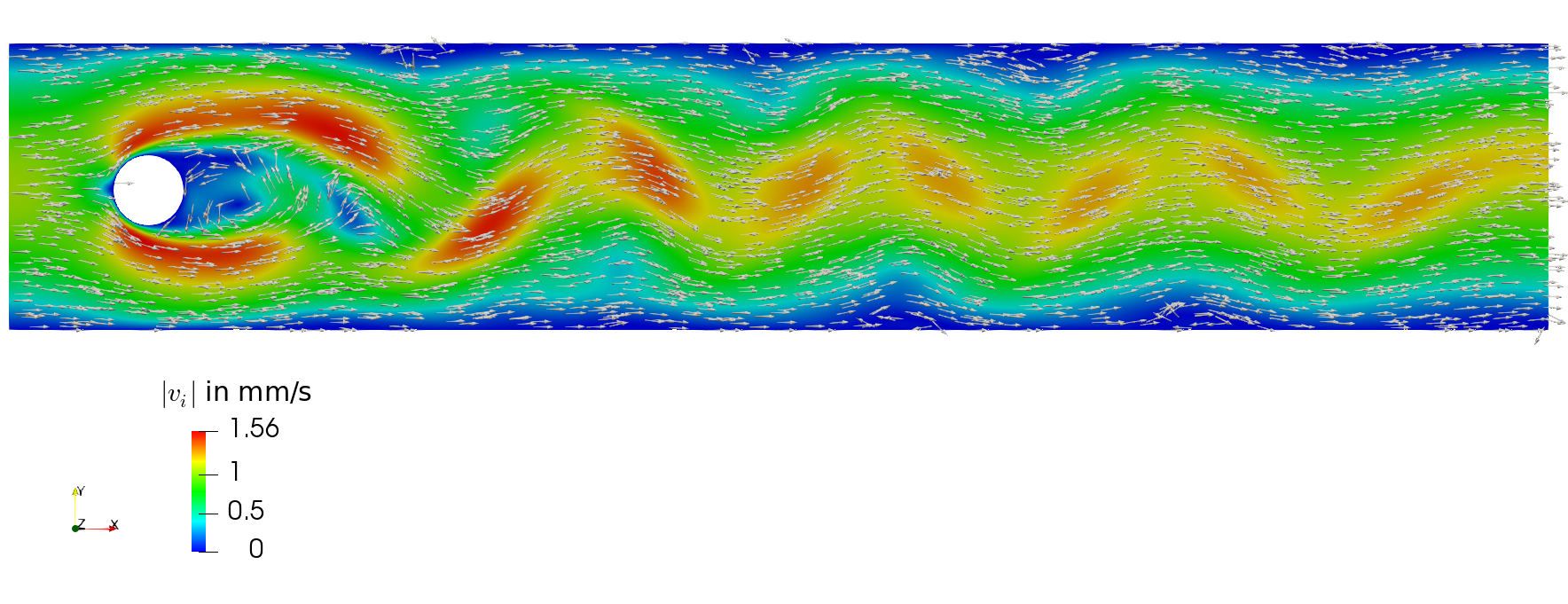}
\vspace{-7mm} \\
\includegraphics[width=0.85\textwidth]{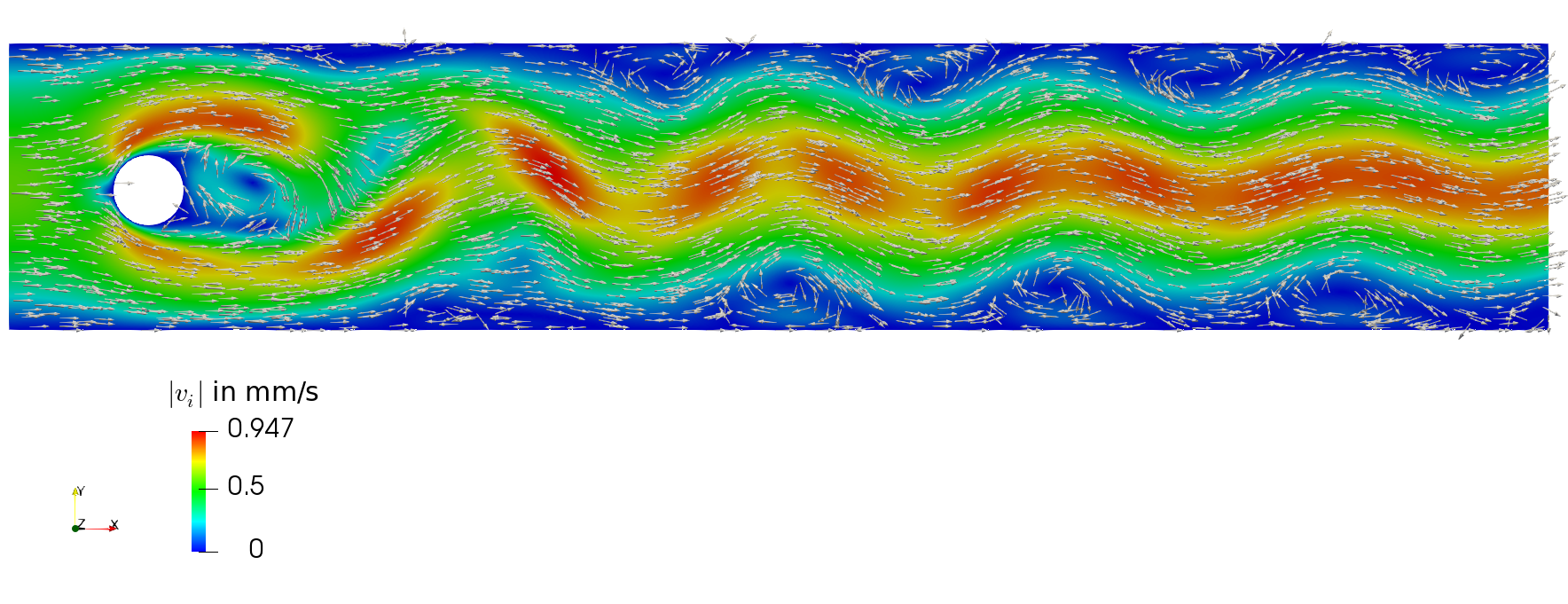}
\vspace{-7mm} \\
\includegraphics[width=0.85\textwidth]{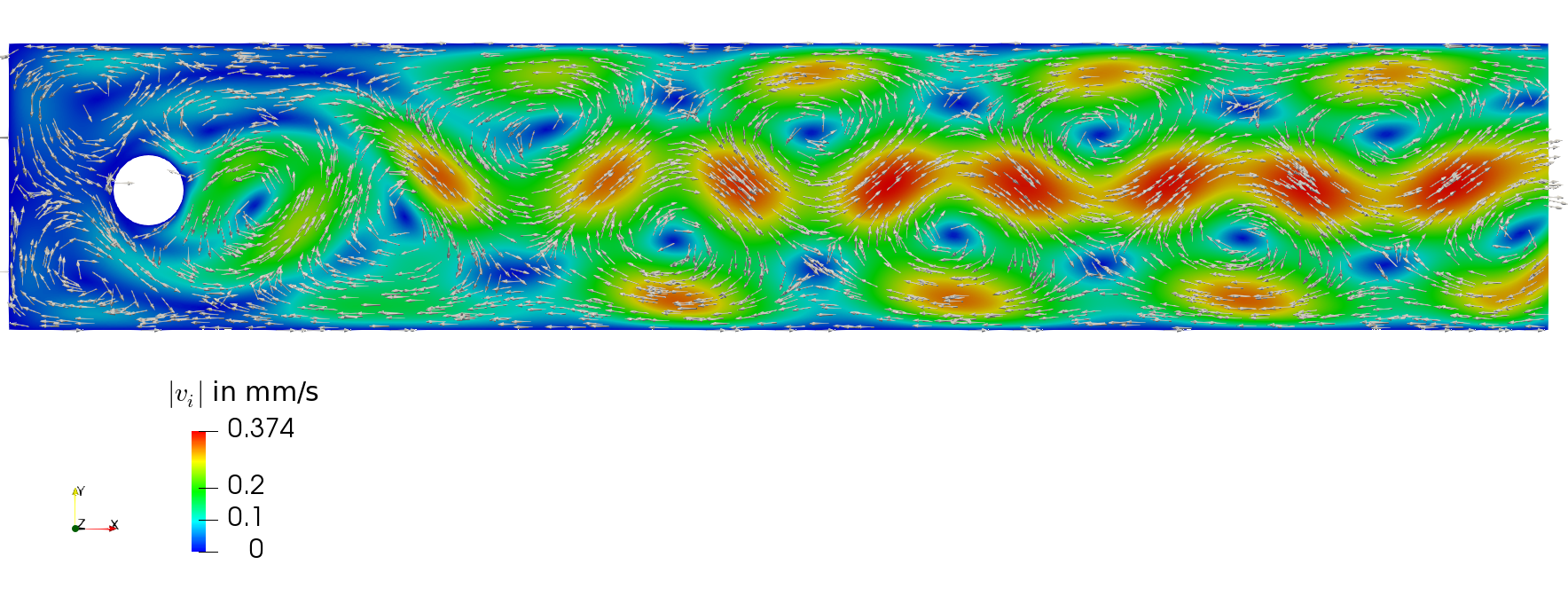}
 \caption{2D computation of the vortex shedding with the configuration as in \cite{john2004reference}, velocity distribution is presented with colors denoting to magnitude and with arrows showing their direction at times 4, 5, 6, 7, 8\,s.}
 \label{fig:over.cyl.mesh2D} 
\end{figure}

A rectangle of $\ell=2.2\,\text{m} \times H=0.41\,\text{m}$ has a circular obstacle of diameter $0.1$\,m with its center located at $(0.2,0.2)$. On the upper and lower walls as well as on the obstacle, fluid adheres to the fixed walls. Fluid is pumped in from the left hand side, the obstacle is placed vertically not at the center such that fluid flowing around it gets perturbed differently on the upper and lower sides of the obstacle. Hence, vortex shedding behind the obstacle generates wakes. On the right end, the \textsc{Dirichlet} boundary condition is set for the pressure $p=p_\Ref$. On the left, where fluid enters the domain, the velocity profile:
\begeq
v_i = \begin{pmatrix} \displaystyle v_\text{max} \sin\Big(\uppi \frac{t}{8}\Big)\frac{4 y (H-y)}{H^2} \\ 0 \end{pmatrix} \ ,
\eqend
is applied with $v_\text{max}=1.5\,\text{m/s}$ leading to the \textsc{Reynolds} number $Re=100$ for fluid of $\rho=1\,\text{kg/m$^3$}$ and $\mu=0.001\,\text{Pa\,s}$. Of course, such a fluid performing an incompressible flow is difficult to find in reality; however, the benchmark problem needs to be seen as a computationally challenging problem demonstrating the strength and robustness of the proposed code since a computation in such a low kinematic viscosity, $\nu=\mu/\rho$, is known to generate numerical problems. By using $\lambda=0.01$\,Pa and $\Delta t=1/1600$\,s, we have successfully computed up to 8\,s such that the velocity increases up to 4\,s and then decreases (sinusoidally, in a half period).

\begin{figure} 
\centering
\includegraphics[width=0.7\textwidth]{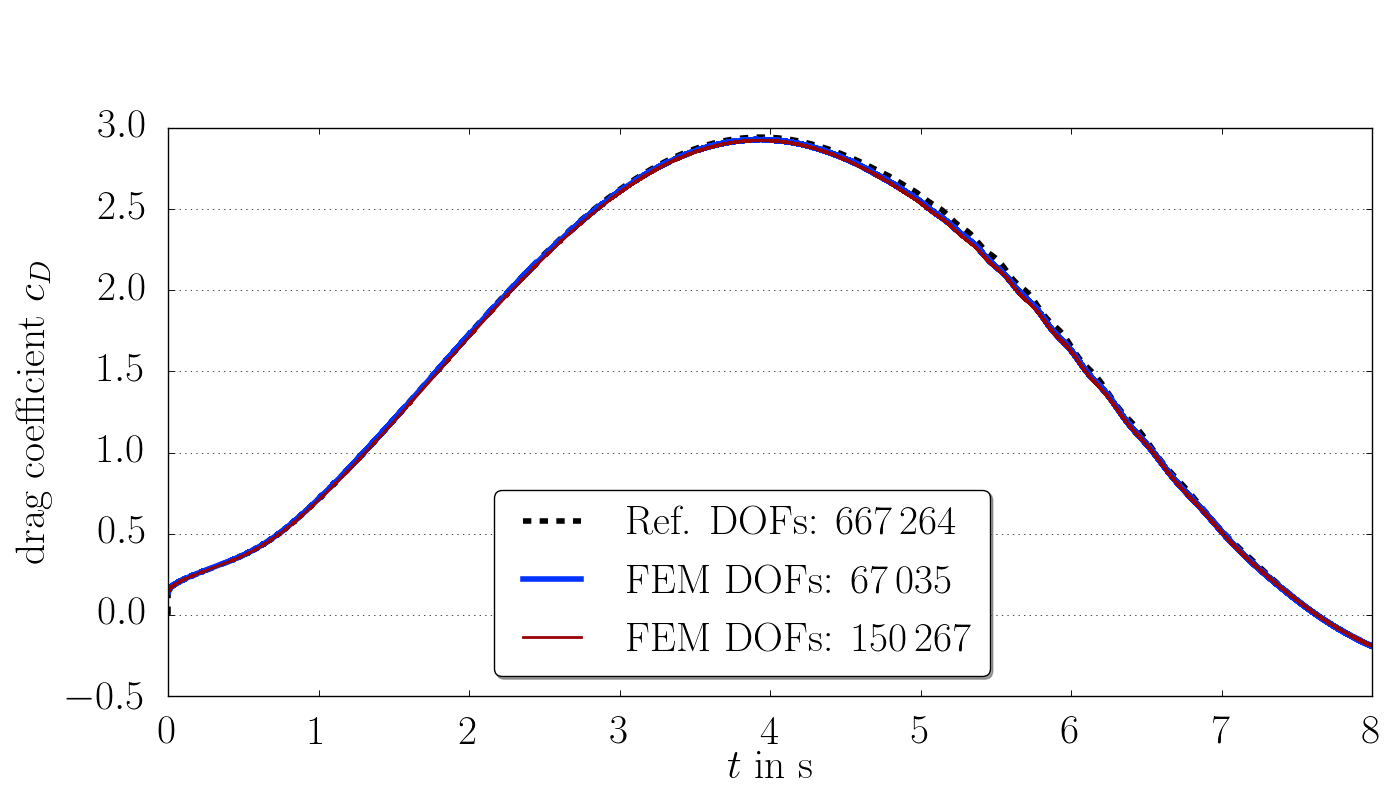}
\includegraphics[width=0.7\textwidth]{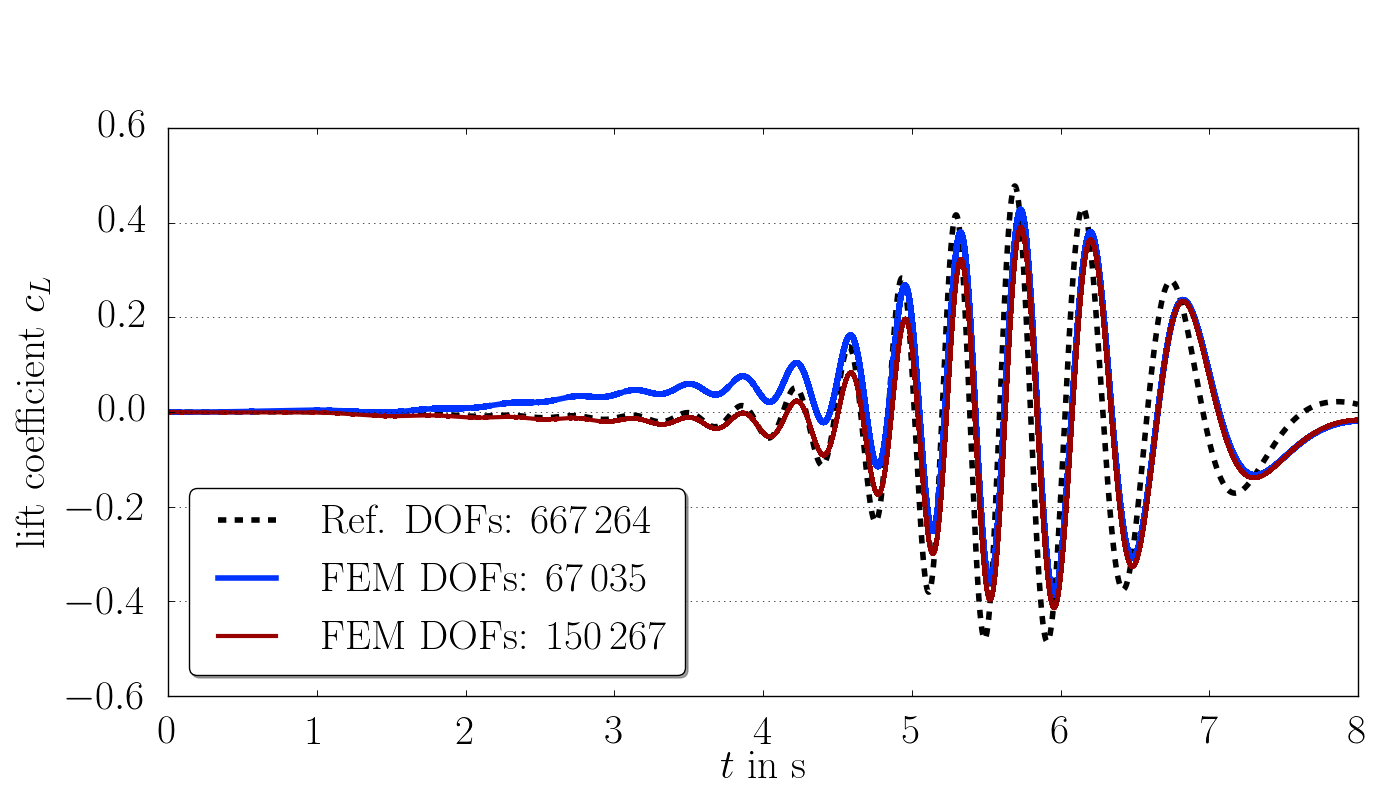}
 \caption{Drag (top) and lift (bottom) coefficients caused by the force applied on the cylinder. FEM denotes the computation presented herein and Ref. indicates the reference solution obtained from \href{http://www.featflow.de/en/benchmarks/cfdbenchmarking/flow/dfg\_benchmark3\_re100.html}{http://www.featflow.de/en/benchmarks/cfdbenchmarking/flow/dfg\_benchmark3\_re100.html} }
 \label{fig:over.cyl.mesh2Dcompare} 
\end{figure}

The vortex shedding behind the obstacle occurs because of the boundary layer separation. This separation is due to the changing pressure gradients on upper and lower parts of the obstacle. There are two stagnation points visible in front of and behind the obstacle. Along the boundary of the obstacle, pressure gradient changes its sign at the top point. This change in pressure gradient resolves a wake behind the obstacle, as the pressure is lower leading to a flow separation. We have modeled this phenomenon with the same element type and size around the boundary, in other words, without any special boundary layer modeling. This benchmark problem is chosen for proving the robustness of the proposed method. In Fig.\,\ref{fig:over.cyl.mesh2Dcompare}, 
a quantitative analysis is shown between the reference solution and the FEM computation by using the drag and lift coefficients:
\begeq
c_D = \frac{-2}{v_\text{mean}^2 \ell} \int_\Om 
\Big( \frac{\mu}{\rho} v_{i,j} v^D_{i,j} 
+ v_i v_{j,i} v^D_j 
- p v^D_{i,i} \Big) \d v \ , \\
c_L = \frac{-2}{v_\text{mean}^2 \ell} \int_\Om 
\Big( \frac{\mu}{\rho} v_{i,j} v^L_{i,j} 
+ v_i v_{j,i} v^L_j 
- p v^L_{i,i} \Big) \d v
\eqend
where $\t v^D = (1,0)$ and $\t v^L = (0,1)$ on the obstacle leading to the forces along the flow direction (drag) and perpendicular to the flow direction (lift). The forces are normalized by the mean velocity, $v_\text{mean}=1$\,mm/s, and characteristic length (diameter of the obstacle), $\ell=0.1$\,mm. The reference solution and the herein proposed method, both solve the same equations by the finite element method. As we have shown the convergence  in the last section, we demonstrate not the quantitative agreement but the reliability of the proposed method. Even with a low amount of degrees of freedom (DOFs), a numerical solution is possible without stabilization and by using the same type of form functions for pressure and velocity. In order to examine the capability of the method, we choose the model with DOFs$=67\,035$ and change the viscosity and inlet velocity to obtain a flow with $Re=1000$ reached in 1\,s and held 100\,s long. Again we managed to compute without numerical problems the analogous \textsc{Karman} vortex street with a different formation as depicted in Fig.\,\ref{fig:over.cyl.mesh2D}.
\begin{figure} 
\centering
\includegraphics[width=0.85\textwidth]{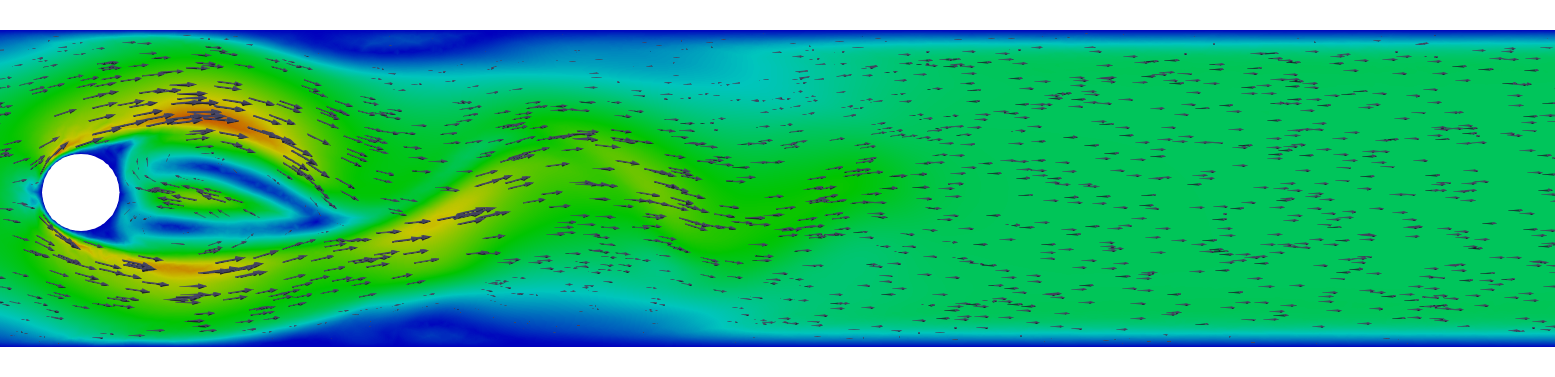}
\vspace{-4mm} \\
\includegraphics[width=0.85\textwidth]{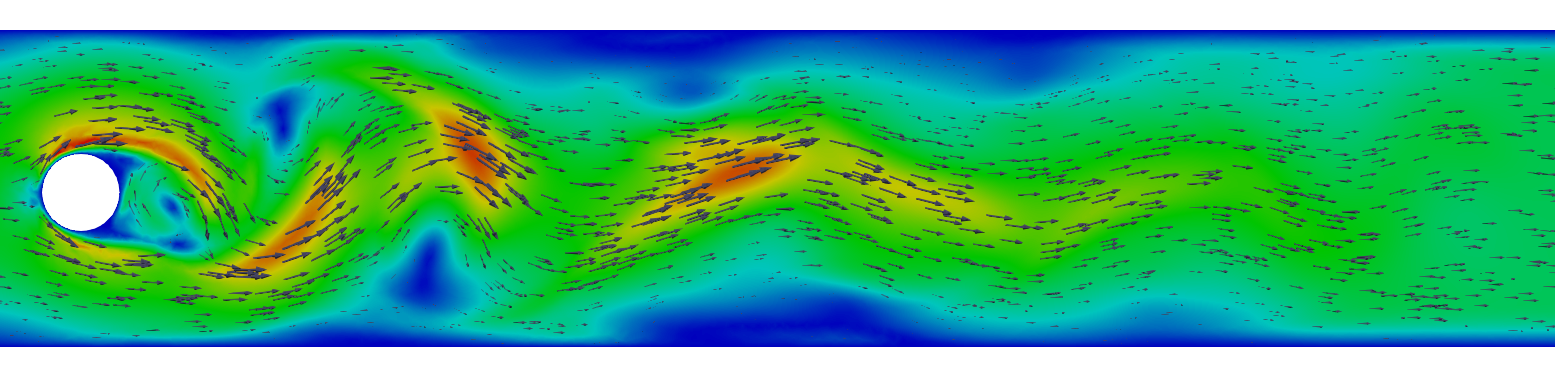}
\vspace{-4mm} \\
\includegraphics[width=0.85\textwidth]{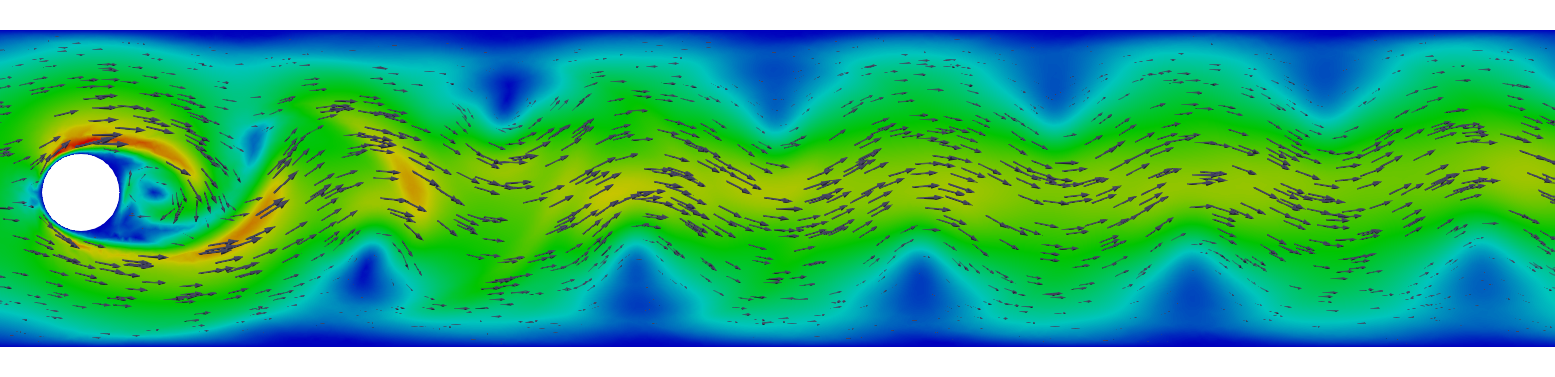}
\vspace{-4mm} \\
\includegraphics[width=0.85\textwidth]{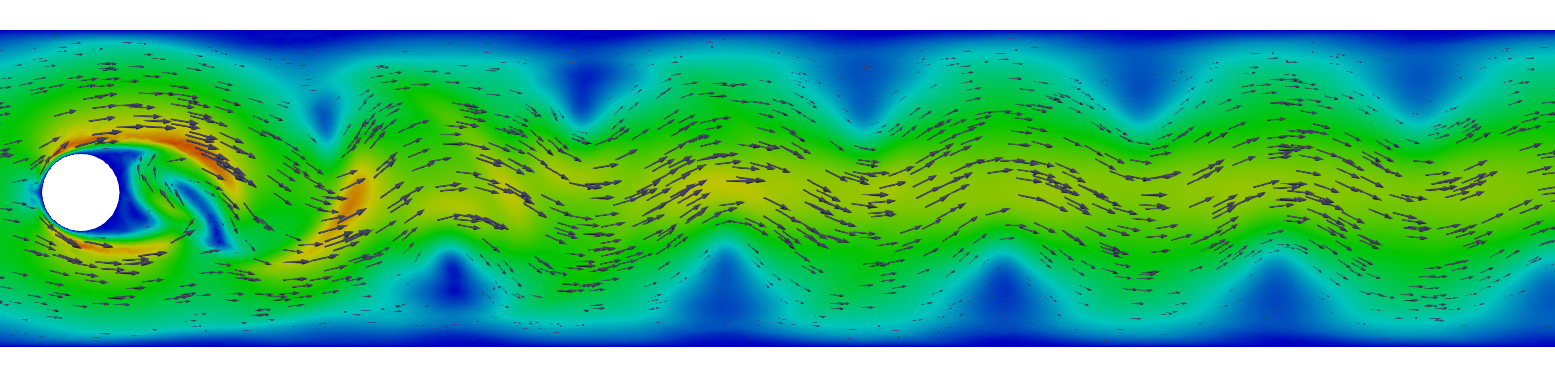}
\vspace{-4mm} \\
\includegraphics[width=0.85\textwidth]{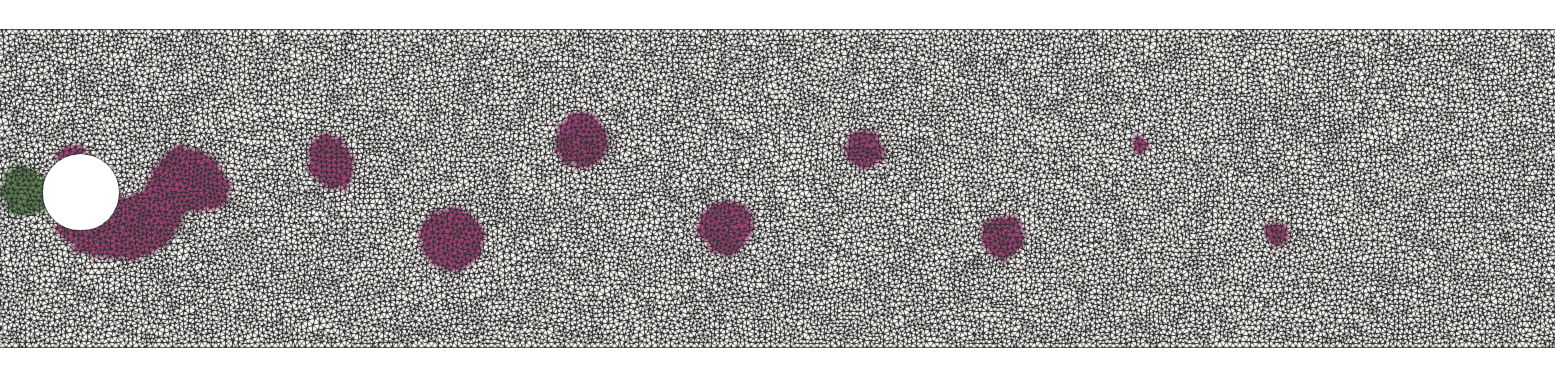}
 \caption{2D computation of the vortex shedding with $Re=1000$ leading to the \textsc{Karman} vortex street, velocity distribution is presented as arrows and their magnitude as colors at $t=15$\,s where the perturbation starts invoking the first vortex, at $t=25$\,s where the vortex street is forming, at $t=65$\,s where the vortex shedding is affecting the whole domain, and at $t=80$\,s where the periodic street is visible and pressure distribution is shown on top of the finite elements discretization with an increased pressure before and moving decreased pressure ``islands'' after the cylinder.}
 \label{fig:over.cyl.mesh2D} 
\end{figure}

\section{Conclusion}

A new method is proposed to compute fluid dynamics of isothermal and incompressible flows by means of the FEM. We have rigorously investigated the convergence and the accuracy of the proposed approach by using closed-form solutions. Convergence in space and time is difficult to achieve in FEM for flow problems. The accuracy is only possible by using extremely high degrees of freedom. With the proposed method, we have attained a good accuracy with relatively coarse meshes. Local monotonic convergence in space and time is a remarkable quality of the FEM and our method exploits this feature. We stress that this property is of paramount importance for problems where the accurate solution is not known. Based on the local monotonic convergence, \textit{a posteriori} error analysis is possible. Hence, the proposed method is reliable. Moreover, we have demonstrated the robustness of the implementation in open-source packages called FEniCS by solving benchmark problems. All used codes are publicly available on the web site in \cite{compreal} to be used under the GNU Public license \cite{gnupublic}. Further research is being conducted for computing real-life problems and verification of the proposed method with the aid of experimental results.

\section*{Acknowledgement}
B. E. Abali had the pleasure to have discussed and worked together with Prof. \"{O}mer Sava\c{s} at the University of California, Berkeley.

\bibliographystyle{plain}
\bibliography{abali_flow}

\end{document}